\documentclass [11pt,a4paper]{article}

\usepackage{mathptmx}
\usepackage{amssymb}
\usepackage{amsmath}
\usepackage{eqnarray}
\usepackage{txfonts}
\usepackage[all]{xy}
\usepackage{multirow}
\usepackage{diagbox}
\def\dse#1{\vskip 0.6cm\noindent
        {\large\bf #1}
        \vskip 0.4cm}

\begin{document}

\begin{center}
\textbf{$\Large${Linear Codes over Galois Ring  $GR(p^2,r)$ Related to Gauss sums} }\footnote {The work of A. Zhang  was supported by National Natural Science Foundation of China(NSFC) under Grant 11401468. 
The work of J. Li was supported by NSFC under Grant 61370089,11501156 and the Anhui Provincial Natural Science Foundation under Grant 1508085SQA198. The work of K. Feng was supported by the NSFC under Grant 11471178,11571007 and the Tsinghua National Lab.for Information Science and Technology. \\
A. Zhang is with the Department of Mathematical Sciences, Xi¡¯an University of Technology, Shanxi, 710048, China(e-mail:zhangaixian1008@126.com )\\
J. Li is with the School of Mathematics, Hefei University of Technology, Anhui, 230001, China(e-mail: lijin\_0102@126.com )\\
K. Feng is with the Department of Mathematical Sciences, Tsinghua University, Beijing, 100195, China(e-mail: kfeng@math.tsinghua.edu.cn)}

\end{center}

\begin{center}
\small Aixian Zhang,  Jin Li,  Keqin Feng
\end{center}


\noindent\textbf{Abstract:} Linear codes over finite rings become one of hot topics in coding
theory after Hommons et al.([4], 1994) discovered that several remarkable nonlinear binary codes
with some linear-like properties are the images of Gray map of linear codes over $Z_4$. In
this paper we consider two series of linear codes $C(G)$ and $\widetilde{C}(G)$ over Galois ring
$R=GR(p^2,r)$, where $G$ is a subgroup of $R^{(s)^*}$ and $R^{(s)}=GR(p^2,rs)$.
We present a general formula on $N_\beta(a)$ in terms of Gauss sums on $R^{(s)}$ for each $a\in R$,
where $N_\beta(a)$ is the number of a-component of the codeword $c_\beta\in C(G) (\beta\in R^{(s)})$ (Theorem 3.1). We have determined the complete Hamming weight distribution of $C(G)$ and the minimum Hamming distance of $\widetilde{C}(G)$ for some special G (Theorem 3.3 and 3.4).
We show a general formula on homogeneous weight of codewords in $C(G)$ and $\widetilde{C}(G)$ (Theorem 4.5)
for the special $G$ given in Theorem 3.4. Finally we obtained series of nonlinear codes over $\mathbb{F}_{q} \ (q=p^r)$ with two Hamming distance by using Gray map (Corollary 4.6).

\noindent\textbf{keywords:} linear code, Galois ring, Gauss sum, homogeneous weight, Gray map, weight distribution.\\

\dse{1~~Introduction}
Linear codes over finite rings, like $R=Z_4$, have been researched from as early as 1970's, but a series of active efforts appears since the paper of Hammons et al.[4]
in 1994 where the authors discovered that several remarkable nonlinear binary codes with some linear-like properties are the images of Gray map of linear codes over $Z_4$.
The underlying finite ring of linear codes has been extended from $Z_4$ to Frobenius rings. Several classification results on linear or cyclic codes over finite rings have been given, the proper(homogeneous) weight
and related Gray map have been found, and the error-correcting ability has been
determined by computing for relatively small parameters or
estimated by using exponential sums over Galois rings[14]. But in general,
to determine the minimal distance and weight distribution of linear codes over Galois ring is a difficult problem.\\

In 1970's, Baumert and McEliece [1,12]
initiated a method to compute the weight distribution of irreducible cyclic codes over finite fields by
using Gauss sums. The method is developed by Langevin, Van Der Vlugt et al.([9, 18]). The main aim of this paper is to extend this
method to Galois ring $R=GR(p^2,r)$. Namely, we compute the weight distribution for a wide class of linear
codes over  $R=GR(p^2,r)$ by using Gauss sums on Galois rings. Then by using the homogeneous weight for linear codes on $R$ and the
Gray map given by $[3,7]$, we obtain a series of good nonlinear codes over finite field $\mathbb{F}_{q}.$\\

 The paper is organized as following. In section 2 we introduce basic facts on
 Galois ring $R=GR(p^2,r)$ including the additive and multiplicative structure of $R$.
 The group $\widehat{R}$ of additive characters and the group $\widehat{R^*}$ of multiplicative characters, and
 Gauss sums on $R$. Section 3 is the main body of our paper where we express the Hamming weight distribution of
 a large class of linear codes over $R$ by Gauss sums. In section 4, we introduce the homogeneous weight for the linear codes over $R$
 and Gray map from $R=GR(p^2,r)$ into $\mathbb{F}_{q}^q \ ( q=p^r )$ . As Gray map's image of linear codes over $R$, we obtain a series of good nonlinear codes
 over finite field $\mathbb{F}_{q}$. At last we make some conclusion remarks in section 5.

\dse{2~~Basic Facts on the Galois ring $R=GR(p^2,r)$}
In this section, we introduce some basic facts on the Galois ring $R=GR(p^2,r)$. FACT(1)-(4) can be seen in Wan's book[16] and for FACT(5) and (6),
we refer to Li et al.[10].\\

 \noindent\textbf{FACT(1)} Let $p$ be a prime number, $Z_{p^2}=\mathbb{{Z}}/$${p^2}\mathbb{Z}$ and $Z_p=\mathbb{F}_p$ be the finite field with $p$
elements, we have the following "module $p$" homomorphism of rings
\begin{equation*}\text{(mod}p): \ Z_{p^2}\longrightarrow Z_p=\mathbb{F}_p, \ a(\text{mod}p^2)\longmapsto\overline{a}=a(\text{mod}p)\end{equation*}
This mapping can be naturally extended as a homomorphism of polynomial rings
\begin{equation*}\text{(mod}p): \ Z_{p^2}[x]\longrightarrow \mathbb{F}_p[x], f(x)=\sum_ia_ix^i\longmapsto{\overline{f}(x)}=\sum_i\overline{a}_ix^i\tag{2.1}\end{equation*}
The Galois ring $GR(p^2,r)$ is defined by the quotient ring \begin{equation*}R=GR(p^2,r)=\frac{Z_{p^2}[x]}{(h(x))}\tag{2.2}\end{equation*}
where  $h(x)$ is a basic primitive polynomial of degree $r$ in $Z_{p^2}[x]$  which means that $\overline{h}(x)$ is a primitive polynomial of degree $r$
in $\mathbb{F}_p[x]$. Then the $''$module $p''$
map (2.1) induces the following homomorphism   of rings
$$\text{(mod}p): R=\frac{\ Z_{p^2}[x]}{(h(x))}\longrightarrow \overline{R}=\frac{\mathbb{F}_p[x]}{(\overline{h}(x))}=\mathbb{F}_q \ (q=p^r)$$
The kernel is the unique maximal ideal $M=pR$, and $R^*=R\backslash M$ is the(multiplicative) group of units. Therefore $R$ is a (commutative)
local ring. $R$ have three ideal only: $R, M=pR$ and $(0)$.\\

 \noindent\textbf{FACT(2)} Let $\xi$ be a root of $h(x)$ in $R$. Then the order of $\xi$ is $q-1$ and $\overline{\xi}$ is a root of $\overline{h}(x)$,
$\mathbb{F}_{p}[\overline{\xi}]=\mathbb{F}_{q}$. By definition (2.2) of $R$, each element $\alpha$ of $R$ can be expressed uniquely as
$$\alpha=c_0+c_1\xi+\cdots+c_{r-1}\xi^{r-1} \ (c_i\in Z_{p^2})$$
and $R$ is a free $Z_{p^2}-$ module with rank $r$. On the other hand, let
$$T^*=\langle\xi\rangle=\{1,\xi,\xi^2,\cdots,\xi^{q-2}\},  T=T^*\cup\{0\}$$
Then $R=T\oplus pT$. Namely, each element $\alpha$ of $R$ can be expressed  uniquely by
$$\alpha=\alpha_1+p\alpha_2 \ (\alpha_1,\alpha_2\in T)$$
we have $M=pR=pT$. Namely, $\alpha\in M$ if and only if $\alpha_1=0$ and
$\alpha\in R^*$ if and only if $\alpha_1\in T^*$. Particularly,$\overline{T}=\mathbb{F}_q,\overline{{T}^*}=\mathbb{F}_q^*$
and $|R|=q^2=p^{2r}, |M|=q, |R^*|=q(q-1)$.\\

 \noindent\textbf{FACT(3)} The group $R^*$ of units has the direct
decomposition
 $$R^*=T^*\times(1+M)$$
 where  $T^*=\langle\xi\rangle$ is a cyclic group with order $q-1$, and the multiplicative group $1+M=1+pT$ is isomorphic to the additive group $\mathbb{F}_q$
 by
\begin{gather*}1+pT\widetilde{\longrightarrow}(\mathbb{F}_q,+),\ \ \
1+pc\longmapsto\overline{c} \ (c\in T) \end{gather*}
Thus $(1+M)$ is an elementary $p-$group. Namely, $1+M$ is a direct product of $r$ copies of cyclic groups with order $p$.\\

\noindent\textbf{FACT(4)} $R/{Z}_{p^2}$
is Galois extension of rings and the Galois group
is the following cyclic group
of order $r$
$$Gal(R/{Z}_{p^2})=\langle\sigma_p\rangle$$
 \noindent where $\sigma_p$ is defined by
$$\sigma_p(\alpha)=\alpha_1^p+p\alpha_2^p \ \ (\alpha=\alpha_1+p\alpha_2,\alpha_1,\alpha_2\in T).$$
Then we have the trace mapping:
$$T_{{Z}_{p^2}}^R:\ R\longrightarrow{Z}_{p^2}, \  \ T_{{Z}_{p^2}}^R(\alpha)=\sum_{i=0}^{r-1}\sigma^i_p(\alpha)=\big(\sum_{i=0}^{r-1}\alpha_1^{p^i}\big)+p\big(\sum_{i=0}^{r-1}\alpha_2^{p^i}\big)$$
This is a surjective homomorphism of ${Z}_{p^2}-$ modules.\\

More general, let $Q=q^s(=p^{rs}),$ and $$R^{(s)}=GR(p^2,rs)=\frac{Z_{p^2}[x]}{(h^{(s)}(x))}=Z_{p^2}[\xi^{(s)}].$$
where $h^{(s)}(x)$ is a basic primitive polynomial in $Z_{p^2}[x]$ with degree $rs$, and $\xi^{(s)}$ is a root
of $h^{(s)}(x)$ in $R^{(s)}$. Then $T^{(s)^*}=\langle\xi^{(s)}\rangle$ is a cyclic group with order $Q-1$,
$T^{(s)}=T^{(s)^*}\cup\{0\}$ and
\begin{align*}R^{(s)}&=\{\alpha_1^{(s)}+p\alpha_2^{(s)}: \alpha_1^{(s)},\alpha_2^{(s)}\in T^{(s)}\}\\
M^{(s)}&=pT^{(s)},\text{ the (unique) maximal ideal of} \ R^{(s)}\\
R^{(s)^*}&=R^{(s)}\backslash M^{(s)}=\{\alpha_1^{(s)}+p\alpha_2^{(s)}: \alpha_1^{(s)}\in T^{(s)^*},\alpha_2^{(s)}\in T^{(s)}\}\\
&=T^{(s)^*}\times(1+M^{(s)})\ \text{(direct product)}\end{align*}
We have the following isomorphism of groups
$$(1+M^{(s)},\cdot)\widetilde{\longrightarrow}(\mathbb{F}_Q,+), \ 1+pc\longmapsto\overline{c}(\text{mod} p) \ (c\in T^{(s)})$$
The extension $R^{(s)}/R$ of rings is Galois extension with Galois group
$$Gal(R^{(s)}/R)=\langle\sigma_q\rangle \ (\text{cyclic group of order} \ s)$$
where
$$\sigma_q(\alpha_1^{(s)}+p\alpha_2^{(s)})={ \alpha_1^{(s)}}^q+p{\alpha_2^{(s)}}^q  \  \ \ (\text{for} \  \alpha_1^{(s)},\alpha_2^{(s)}\in T^{(s)})$$
Then we have the trace mapping
$$T_R^{R^{(s)}}: R^{(s)}\longrightarrow R, \ \ T_R^{R^{(s)}}(\alpha)=\sum_{i=0}^{s-1}\sigma_q^i(\alpha)\  \ \ (\text{for} \  \alpha\in R^{(s)})$$
$T_R^{R^{(s)}}$ is a surjective homomorphism of $R-$ modules and the following diagram is commutative\\
\begin{equation*}\xymatrix{
 R^{(s)}\ar[d]_{ T_R^{R^{(s)}}} \ar[r]^{(mod p)} & \mathbb{F}_Q \ar[d]^{{T}_q^Q} \\
  R \ar[d]_{T^R_{Z_{p^2}}} \ar[r]^{(mod p)} & \mathbb{F}_q \ar[d]^{{T}_p^q} \\
  Z_{p^2} \ar[r]^{(mod p)} &\mathbb{F}_p   }\tag{2.3}\end{equation*}
where ${{T}}_q^Q$ is the trace mapping from $\mathbb{F}_Q$ to $\mathbb{F}_q$.\\

\noindent\textbf{FACT (5)} The additive characters and  multiplicative  characters  of $R=GR(p^2,r)$.\\

 A character of the additive group $(R,+)$ is called
an additive character of $R$. The group of all additive characters of
$R$ is
$$\widehat{R}=\{\lambda_\beta:\beta\in R\},$$
where (for each positive integer $m,$ let $\zeta_m=e^{\frac{2\pi\sqrt{-1}}{m}}\in\mathbb{C}$)\\
 $$\lambda_\beta:\ R\longrightarrow\langle\zeta_{p^2}\rangle, \  \lambda_\beta(x)=\zeta_{p^2}^{T_{Z_{p^2}}^R(\beta x)} \ (x\in R)$$

A  character of the unit group $(R^*,\cdot)$ is called a  multiplicative character of $R$.

\noindent Since $R^*=T^*\times(1+M)$ and
$1+M\cong(\mathbb{F}_q,+)$, we
know that the group of multiplicative characters of $R$ is
 \begin{align*}\widehat{R}^*&=\widehat{T}^*\times{(1+M)}^{\widehat{}}\cong\widehat{T}^*\times{(\mathbb{F}_q,+)}^{\widehat{}}\ \ (T^*=\langle\xi\rangle, \ M=pT)\\
&=\{\omega^i\varphi_b:  \ 0 \leq i \leq q-2,\ b\in T\},\end{align*}
where
$$\omega(1+M)=1,  \ \    \omega(\xi)=\zeta_{q-1};$$
$$\varphi_b(T^*)=1, \ \   \varphi_b(1+px)=\zeta_p^{{T}_p^q(\overline{bx})}\ \   for  \  x\in T.$$

\noindent\textbf{FACT (6)} Gauss sums on R\\

The calculation of Gauss sums on general finite rings is initiated by Lamprecht [8] in 1989.
The Gauss sums on Galois rings has been computed in [13,7,10]. In this paper we restate the result in [10].\\

 Let $\chi$ and $\lambda$
be a multiplicative and additive character respectively. The Gauss sum on $R$ is defined by
 \begin{equation*}G(\chi,\lambda)=G_R(\chi,\lambda)=\sum_{x\in
R^*}\chi(x)\lambda(x)\in\mathbb{Z}[\zeta_{(q-1)p^2}].\end{equation*}
The following result shows that the non-trivial values of all Gauss
sums $G(\chi,\lambda)$ on $R$ are essentially reduced into ones of the
Gauss sums on finite field $\mathbb{F}_q$. For the Gauss sums on finite fields we refer to Lidl and Niederreiter's book [11].

\noindent{\bf Lemma~2.1[10]}  For $\chi=\omega^i\varphi_b\in \widehat{R}^* (0 \leq i \leq q-2,\ b\in T)$ and $\lambda=\lambda_\beta\in \widehat{R} \ (\beta\in R)$, we have\\
(\uppercase\expandafter{\romannumeral 1})(trivial case)\begin{eqnarray*}G(\chi,\lambda)=\begin{cases}
q(q-1), &if \  \chi=1 \ and  \ \lambda=1 \cr 0, &if \ \chi\neq1 \ and \ \lambda=1\cr -q,
&if \ \chi=1 \ and  \lambda=\lambda_\beta, \ (\beta\in M\backslash\{0\})\cr 0,
&if \ \chi=1 \ and  \lambda=\lambda_\beta,  \  (\beta\in R^*)
\end{cases}\end{eqnarray*}
(\uppercase\expandafter{\romannumeral 2}) Let $\chi\neq 1((i,b)\neq(0,0)\in Z_{q-1}\times T)$ and $\lambda=\lambda_\beta\neq1 (\beta\neq0)$,\\

\noindent(\uppercase\expandafter{\romannumeral 2}.1) If $\beta\in R^*$, then $G(\chi,\lambda_\beta)=\overline{\chi}(\beta)G(\chi)$, where

$$G(\chi)=G(\chi,\lambda_1)=\sum_{x\in R^*}\chi(x)\zeta_{p^2}^{T_{Z_{p^2}}^R(x)}$$

 If $\beta=py, y\in T^*$, then $G(\chi,\lambda_\beta)=\overline{\chi}(y)G(\chi,\lambda_p)$, where

 $$G(\chi,\lambda_p)=\sum_{x\in R^*}\chi(x)\zeta_{p^2}^{T_{Z_{p^2}}^R(py)}=\sum_{x\in R^*}\chi(x)\zeta_p^{{T}_p^q(\overline{x})}$$
\noindent(\uppercase\expandafter{\romannumeral 2}.2) For  $\chi=\omega^i\varphi_b\neq 1((i,b)\neq(0,0)\in Z_{q-1}\times T)$, then
\begin{equation*}\hspace{2em}G(\chi)=\begin{cases}
0, &if \ \chi(1+M)=1 (\Leftrightarrow b=0)  \cr {q\omega^i(b')\zeta_{p^2}^{T^R_{Z_{p^2}}(b')}}, & if
\ \chi(1+M)\neq1 (\Leftrightarrow b\in T^*)\end{cases} \tag{2.4} \end{equation*}
where  $b'=b$ for $p=2$ and $b'=-b$ for $p\geq3,$ and
\begin{equation*}\hspace{3em}G(\chi,\lambda_p)=\begin{cases}
 qG_q(\omega^i), &if \ \chi(1+M)=1 (\Leftrightarrow b=0)  \cr  0, & if \  \chi(1+M)\neq1 (\Leftrightarrow b\in T^*) \end{cases}\tag{2.5}\end{equation*}
where $G_q(\omega^i)=\sum\limits_{ \ x\in
\mathbb{F}_q^*}\omega^i(x)\zeta_p^{{T}_p^q({x})}$ is the
Gauss sum on finite field  $\mathbb{F}_q$.\\

 This result has been proved in [10]. Here we reproduce the proof for reader's convenience.\\

(\uppercase\expandafter{\romannumeral 1}) The case $\lambda=1$ is easy. Now we assume that $\chi=1$ and $\lambda=\lambda_\beta$ for $\beta\in R\backslash\{0\}$. In this case,
$$G(\chi,\lambda)=\sum_{x\in R^*}\lambda(x)=-\sum_{x\in M}\lambda(x)=-\sum_{y\in T}\zeta_{p^2}^{T_{Z_{p^2}}^R(py\beta)}=-\sum_{y\in\mathbb{F}_q}\zeta_p^{{T}_p^q(y\overline{\beta})}$$
If $\beta\in R^*$, then $\overline{\beta}\in\mathbb{F}_q^*$ so that the right-hand side is 0. If $\beta\in M\backslash\{0\}$, then $\overline{\beta}=0$ and the right-hand side is $-q$.\\

(\uppercase\expandafter{\romannumeral 2}) Proof of (\uppercase\expandafter{\romannumeral 2}.1) is easy. For (\uppercase\expandafter{\romannumeral 2}.2), we denote $Tr=T_{Z_{p^2}}^R$ and $\overline{T}r={T}_p^q$, then
\begin{align*}G(\chi)&=\sum\limits_{ \ x_1\in Tr^*\atop x_2\in
T}\omega^i(x_1)\varphi_b(1+px_2)\zeta_{p^2}^{Tr(x_1(1+px_2))}\\
&=\sum_{ x_1\in T^*}\omega^i(x_1)\zeta_{p^2}^{Tr(x_1)}\sum_{x_2\in T}\zeta_{p}^{\overline{T}r(\overline{x}_2(\overline{b}+\overline{x}_1))}\\
&=q\sum\limits_{ \ x_1\in T^*\atop \overline{x}_1=-\overline{b}
}\omega^i(x_1)\zeta_{p^2}^{Tr(x_1)}
\end{align*}
If $b=0$, there is no $x_1\in T^*$ such that $\overline{x}_1=0$, so that $G(\chi)=0$. If $b\in T^*$, there is unique $b'\in T^*$ such that $\overline{b'}=-\overline{b}$, so that $G(\chi)=q\omega^i(b')\zeta_{p^2}^{Tr(b')}$. For $p\geq 3, \ -b\in T^*$ so that $b'=-b$. For $p=2$, we have $-\overline{b}=\overline{b}$, so that $b'=b$. We completes the proof of (2.4). For the Gauss sum $G(\chi,\lambda_p)$, we have
$$G(\chi,\lambda_p)=\sum\limits_{ \ x_1\in T^*\atop x_2\in
T}\omega^i(x_1)\varphi_b(1+px_2)\zeta_{p}^{\overline{T}r(\overline{x}_1)}=G_q(\omega^i)\sum_{x_2\in T}\varphi_b(1+px_2)$$
Since $1+M=1+pT$ we get
\begin{eqnarray*}\hspace{3em}\sum_{x_2\in T}\varphi_b(1+px_2)=\sum_{x_2\in T}\zeta_p^{\overline{T}r(\overline{b}\overline{x}_2)}=\begin{cases}
 q, &if \  b=0  \cr  0, & if \   b\in T^* \end{cases}\end{eqnarray*}
This completes the proof of (2.5).\\

\noindent\textbf{{3. Linear Codes $C=C(G)$ over $R=GR(p^2,r)$}}\\

In this section we consider a class of linear codes $C=C(G)$ over Galois ring $R=GR(p^2,r)$, where
$G$ is a subgroup of $R^*$. We computer the weight distribution of $C$ by using Gauss sums on Galois ring given in section 2.\\

Let $Q=q^s, q=p^r$ where $p$ is a prime number. For Galois ring $R=GR(p^2,r)$
 and $R^{(s)}=GR(p^2,sr)$, their unit groups are
 $$R^{(s)^*}=T^{(s)^*}\times(1+pT^{(s)}), \ R^*=T^*\times(1+pT)$$
where
$$T^{(s)^*}=\langle\xi^{(s)}\rangle, \ T^{(s)}=T^{(s)^*}\cup\{0\}$$
$$T^*=\langle\xi\rangle, \ T=T^*\cup\{0\},  \ \text{and} \ \xi=\xi^{(s)^{\frac{Q-1}{q-1}}}.$$
Each subgroup $G$ of $R^{(s)^*}$ has the following structure
$$G=D\times(1+pV)$$
where $D=\langle\xi^{(s)^e}\rangle$ is a subgroup of $T^{(s)^*}$, $Q-1=ef$, and $1+pV$ is a
subgroup of $1+pT^{(s)}$. Since $1+pT^{(s)}$ is an elementary $p-$group with rank $rs$, $1+pV$
is an elementary $p-$group with rank $d$ $(0\leq d\leq rs)$ and $\overline{V}$ is a $d-$dimensional
$\mathbb{F}_{p}-$subspace of $\mathbb{F}_Q=\overline{T^{(s)}}.$ Thus the size of $G$ is
$$n=|G|=fp^d.$$
Let $G=\{x_1,\cdots,x_n\}$, we consider the following code $C=C(G)$ over $R$,
\begin{equation*}C=\{c_\beta=(T_R^{R^{(s)}}(\beta x_1),\cdots,T_R^{R^{(s)}}( \beta x_n))\in R^n: \beta\in R^{(s)}\}\tag{3.1}\end{equation*}
This is a $R-$linear code. Namely, $C$ is a $R-$submodule of $R^n$. For $\beta\in R^{(s)}\backslash\{0\}$ and $a\in R$, let $N_\beta(a)$ be the number of $a$-components of the codeword $c_\beta$
\begin{equation*}
N_\beta(a)=\sharp\{1\leq i\leq n: \ T_R^{R^{(s)}}( \beta x_i)=a\}\tag{3.2}\end{equation*}
Let $R=\{a_1,a_2,\cdots,a_{q^2}\}$. For $m_1,\cdots,m_{q^2}\geq 0$, let
$$N(m_1,\cdots,m_{q^2})=\sharp\{\beta\in R^{(s)}: N_\beta(a_i)=m_i \ ( 1\leq i\leq q^2)\}$$
Then $\{N(m_1,\cdots,m_{q^2}): m_1,\cdots,m_{q^2}\geq 0, m_1+\cdots+m_{q^2}=n\}$ is called the complete weight distribution of $C(G)$, and $
{A_0,A_1,\cdots,A_n}$ is called the Hamming weight distribution of $C(G)$, where
$$A_i=\sharp\{\beta\in R^{(s)}: w_H(c_\beta)=i\}$$
and
$$w_H(c_\beta)=\sum_{a\in R\backslash\{0\}}N_\beta(a)=n-N_\beta(0)$$
is the Hamming weight of the codeword $c_\beta$. The number $\{A_0=1,A_1,\cdots,A_n\}$
is called the weight distribution of $C$.\\

Our main result in this section is the following theorem where $N_\beta(a)$ can be expressed
in terms of Gauss sums.\\

\noindent\textbf{Theorem 3.1} Let $C$ be the linear code over $R$ defined by (3.1). For $\beta\in R^{(s)}\backslash\{0\}$ and $a\in R$, $N_\beta(a)$ is defined by (3.2). Then\\

\noindent(1) For $\beta \in M^{(s)}\backslash\{0\}, \ \beta=pb, \ \ \ b\in T^{(s)^*},$\\

(1.1) If $a\in R^*$, then  $N_\beta(a)=0$.\\

(1.2) If  $a\in M\backslash\{0\} \ (M=pT), \ a=pc, \ c\in T^*.$ Then

\begin{equation*}N_\beta(a)=\frac{n}{q}+\frac{n}{q(Q-1)}\sum_{\chi\in(\mathbb{F}_Q^*/\langle^
{\overline{\xi}^{(s)^e}}\rangle)^{\widehat{}}}
\chi(\overline{c}/\overline{b})\overline{G_{q}(\chi)}G_Q(\chi)\tag{3.3}\end{equation*}

\noindent where $G_q$ and $G_Q$ are Gauss sums on $\mathbb{F}_q$ and $\mathbb{F}_Q$ respectively.\\

(1.3)\begin{equation*}N_\beta(0)=\frac{n}{q}+\frac{n(q-1)}{q(Q-1)}\sum_{\chi\in(\mathbb{F}_Q^*/\langle^
{\overline{\xi}^{(s)^{e'}}}\rangle)^{\widehat{}}}\overline{\chi}({\overline{b}})G_Q(\chi)\tag{3.4}\end{equation*}
where $e'=gcd(e,\frac{Q-1}{q-1})$.\\

\noindent(2) For  $\beta\in R^{(s)^*},$  $\beta=\beta_1(1+p\beta_2)$, $\beta_1\in T^{(s)^*}$, $\beta_2\in T^{(s)}$. Then for  $a\in R,$
$$N_\beta(a)=\frac{n}{q^2}+\frac{n}{q^2Q(Q-1)}((\uppercase\expandafter{\romannumeral 1})_a+(\uppercase\expandafter{\romannumeral 2})_a )$$
where
\begin{eqnarray*}(\uppercase\expandafter{\romannumeral 1} )_a=\begin{cases}
 \sum_{\chi\in (R^{(s)^*}/G)^{\widehat{}}}\chi(a/\beta)G_{R^{(s)}}(\chi)\overline{G_R(\chi)}, &for \  a\in R^*  \cr q\sum_{\chi\in (R^{(s)^*}/G(1+M))^{\widehat{}}}\chi(a_2/\beta)G_{R^{(s)}}(\chi)\overline{G_q(\chi)}, &for \  a=pa_2 (a_2\in T^*) \cr q(q-1)\sum_{\chi\in (R^{(s)^*}/GR^*)^{\widehat{}}}\overline{\chi}(\beta)G_{R^{(s)}}(\chi), &for \  a=0 \end{cases}\end{eqnarray*}
\begin{eqnarray*}(\uppercase\expandafter{\romannumeral 2})_a =\begin{cases}
 Q\sum_{\chi\in (\mathbb{F}_Q^*/\langle^
{\overline{\xi}^{(s)^e}}\rangle)^{\widehat{}}}\chi(\overline{a}_1/\overline{\beta}_1)G_Q(\chi)\overline{G_q(\chi)}, &for \  a=a_1+pa_2 \ (a_1\in T^*, a_2\in T)  \cr Q(q-1)\sum_{\chi\in (\mathbb{F}_Q^*/\langle^
{\overline{\xi}^{(s)^{e'}}}\rangle)^{\widehat{}}} \ \chi(1/\overline{\beta}_1)G_Q(\chi), &for \  a\in M(=pT)\end{cases}\end{eqnarray*}
where $e'=gcd(e,\frac{Q-1}{q-1})$.\\

(3) Let $M_1=\max \limits_{ \atop b\in\mathbb{F}_Q^*}\big\{\frac{1}{Q-1}\sum_{\chi\in(\mathbb{F}_Q^*/\langle^
{\overline{\xi}^{(s)^{e'}}}\rangle)^{\widehat{}}} \ \chi(b)G_Q(\chi)\}$,

$$M_2=\max \limits_{ \atop \beta=\beta_1+p\beta_2\in R^{(s)^*}}\{\frac{q}{(q+1)Q(Q-1)}\sum_{\chi\in (R^{(s)^*}/GR^*)^{\widehat{}}} \ \chi(\frac{1}{\beta})G_{R^{(s)}}(\chi)+\frac{1}{(q+1)(Q-1)}\sum_{\chi\in(\mathbb{F}_Q^*/\langle^
{\overline{\xi}^{(s)^{e'}}}\rangle)^{\widehat{}}} \ \chi(1/\overline{\beta}_1)G_Q(\chi)\}$$

If $M_1< 1$ and $M_2<1$, then the size of the code $C$ is $|C|=Q^2$ and the minimum Hamming distance is
$$d_H(C)=min\{\frac{n(q-1)}{q}(1-M_1), \frac{n(q^2-1)}{q^2}(1-M_2)\}$$

\noindent\textbf{Proof} (1) For $\beta=pb$, $ b\in T^{(s)^*}$, $T_R^{R^{(s)}}(\beta x)=pT_R^{R^{(s)^*}}(bx)\in M.$ Therefore
$N_\beta(a)=0$ for $a\in R^*$. Now let $a=pc$, $c\in T^*$. Then

\begin{align*}N_\beta(a)&=\sum\limits_{ \ x\in G\atop T_R^{R^{(s)}}( \beta x)=a
}1=\frac{1}{|R|}\sum_{x\in G}\sum_{\lambda\in \widehat{R}}\lambda(T_R^{R^{(s)}}(\beta x)-a)\\
&=\frac{1}{|R|}\sum_{x\in G}\sum_{A,B\in T}\zeta_{p^2}^{T_{Z_{P^2}}^R[(A+pB)(T_R^{R^{(s)}}(pbx) -pc)]}\\&=\frac{|T|}{|R|}\sum_{x\in G}\sum_{A\in T}\zeta_p^{T_p^q(-\overline{cA})+T_p^Q(\overline{Abx})}
\\&=\frac{|T||G|}{|R|}+\frac{|T||V|}{|R|}\sum_{A\in T^*}\overline{\zeta}_p^{T_p^q(\overline{cA})}\sum_{x\in D}\zeta_p^{T_p^Q(\overline{Abx})}
\end{align*}

The last summation is

\begin{align*}\sum_{x\in D}\zeta_p^{T_p^Q(\overline{Abx})}&=\frac{1}{|\frac{T^{(s)^*}}{D}|}\sum_{x\in T^{(s)^*}}\zeta_p^{T_p^Q(\overline{Abx})}\sum_{\chi\in(\frac{T^{(s)^*}}{D})^{\widehat{}}}\chi(x)\\
&=\frac{|D|}{Q-1}\sum_{\chi\in(\mathbb{F}_Q^*/\langle^
{\overline{\xi}^{(s)^{e}}}\rangle)^{\widehat{}}} \ \sum_{x\in\mathbb{F}_Q^*}\chi(x)\zeta_p^{T_p^Q(\overline{Abx})}
\end{align*}
Therefore

\begin{align*}N_\beta(a)&=\frac{n}{q}+\frac{n}{q(Q-1)}\sum_{\chi\in(\mathbb{F}_Q^*/\langle^
{\overline{\xi}^{(s)^{e}}}\rangle)^{\widehat{}}} \ \sum_{A\in \mathbb{F}_q^*}\overline{\zeta}_p^{T_p^q(\overline{c}A)}\overline{\chi}(A\overline{b})G_Q(\chi) \tag{3.5}\\&=\frac{n}{q}+\frac{n}{q(Q-1)}\sum_{\chi\in(\mathbb{F}_Q^*/\langle^
{\overline{\xi}^{(s)^{e}}}\rangle)^{\widehat{}}} \chi(\overline{c}/\overline{b})\overline{G_q(\chi)}G_Q(\chi)
\end{align*}
For $a=0$, by taking $c=0$ in (3.5) we get

\begin{align*}N_\beta(0)&=\frac{n}{q}+\frac{n}{q(Q-1)}\sum_{\chi\in(\mathbb{F}_Q^*/\langle^
{\overline{\xi}^{(s)^{e}}}\rangle)^{\widehat{}}} \overline{\chi}(\overline{b})G_Q(\chi) \sum_{A\in \mathbb{F}_q^*}\overline{\chi}(A) \\&=\frac{n}{q}+\frac{n(q-1)}{q(Q-1)}\sum_{\chi\in(\mathbb{F}_Q^*/\langle^
{\overline{\xi}^{(s)^{e}}}\rangle\mathbb{F}_q^*)^{\widehat{}}} \overline{\chi}(\overline{b})G_Q(\chi)
\end{align*}
Since $\mathbb{F}_q^*=\langle\overline{\zeta}^{(s)\frac{Q-1}{q-1}}\rangle$ we get $\langle\overline{\zeta}^{(s)^e}\rangle\mathbb{F}_q^*=\langle\overline{\zeta}^{(s)^{e'}}\rangle$
where $e'=gcd(e,\frac{Q-1}{q-1})$. Thus we get (3.4).\\

\noindent (2) Suppose that $\beta\in R^{(s)^*}$, then $\beta=\beta_1(1+p\beta_2)$, $\beta_1\in  T^{(s)^*}$ and $\beta_2\in T^{(s)}$. For $a\in R$ we have

\begin{align*}N_\beta(a)&=\frac{1}{|R|}\sum_{x\in G}\sum_{\lambda\in \widehat{R}}\lambda(T_R^{R^{(s)}}(\beta x)-a)\\&=\frac{n}{|R|}+\frac{1}{|R|}\sum_{1\neq\lambda\in \widehat{R}}\lambda(-a)\sum_{x\in G}\lambda(T_R^{R^{(s)}}(\beta x))\\&=\frac{n}{q^2}+\frac{1}{q^2}\sum_{c\in R\backslash\{0\}}\lambda_c(-a)\sum_{x\in G}\lambda^{(s)}_c(\beta x)
\end{align*}
where $\lambda^{(s)}_c\in\widehat{R^{(s)}}$ is defined by $\lambda^{(s)}_c(x)=\zeta_{p^2}^{T^{R^{(s)}}_{Z_{p^2}}(cx)}$ for $x\in R^{(s)}$
and $\lambda_c\in \widehat{R}$ is defined by $\lambda_c(x)=\zeta_{p^2}^{T^{R}_{Z_{p^2}}(cx)}$
for $x\in R$. Then

\begin{align*}N_\beta(a)&=\frac{n}{q^2}+\frac{1}{q^2}\frac{|G|}{|R^{(s)^*}|}\sum_{c\in R\backslash\{0\}}\overline{\lambda}_c(a)\sum_{x\in R^{(s)^*}}\lambda^{(s)}_c(\beta x)\sum_{\chi\in(R^{(s)^*}/G)^{\widehat{}}}\chi(x)\\&=\frac{n}{q^2}+\frac{n}{q^2Q(Q-1)}\sum_{c\in R\backslash\{0\}}\overline{\lambda}_c(a)\sum_{\chi\in(R^{(s)^*}/G)^{\widehat{}}}\overline{\chi}(\beta )G_{R^{(s)}}(\chi,\lambda_c^{(s)})\\&=\frac{n}{q^2}+\frac{n}{q^2Q(Q-1)}((\uppercase\expandafter{\romannumeral 1})_a+(\uppercase\expandafter{\romannumeral 2})_a )\tag{3.6}
\end{align*}
where
\begin{align*}(\uppercase\expandafter{\romannumeral 1})_a&=\sum_{c\in R^*}\overline{\lambda}_c(a)\sum_{\chi\in(R^{(s)^*}/G)^{\widehat{}}}
\overline{\chi}(\beta)G_{R^{(s)}}(\chi,\lambda_c^{(s)})\\&=\sum_{\chi\in(R^{(s)^*}/G)^{\widehat{}}}\overline{\chi}(\beta)G_{R^{(s)}}(\chi)\sum_{c\in R^*}\overline{\lambda}_a(c)\overline{\chi}(c)\\&=\sum_{\chi\in(R^{(s)^*}/G)^{\widehat{}}}
\overline{\chi}(\beta)G_{R^{(s)}}(\chi)\overline{G_R({\chi},{\lambda}_a)} \tag{3.7}
\end{align*}
and
\begin{align*}(\uppercase\expandafter{\romannumeral 2})_a&=\sum_{c\in T^*}\overline{\lambda}_{pc}(a)\sum_{\chi\in(R^{(s)^*}/G)^{\widehat{}}}
\overline{\chi}(\beta)G_{R^{(s)}}(\chi,\lambda_{pc}^{(s)})\\&=\sum_{c\in T^*}\overline{\lambda}_{pc}(a_1)\sum_{\chi\in(T^{(s)^*}/D)^{\widehat{}}}
\overline{\chi}(\beta_1)\overline{\chi}(c)G_{R^{(s)}}(\chi,\lambda_{p}^{(s)})\\
&(by  \ (2.5)\ where \ a=a_1+pa_2, a_1,a_2\in T)\\
&=\sum_{\chi\in(T^{(s)^*}/D)^{\widehat{}}}
\overline{\chi}(\beta_1)G_{R^{(s)}}(\chi,\lambda_{p}^{(s)})\sum_{c\in T^*}\overline{\lambda}_p(ca_1)\overline{\chi}(c)
\\&=Q\sum_{\chi\in(\mathbb{F}_Q^*/\overline{D})^{\widehat{}}}
\overline{\chi}(\overline{\beta}_1)G_Q(\chi)\overline{G_q(\chi,\lambda_{\overline{a}_1})}\tag{3.8}
\end{align*}
(2.1)For $a=a_1+pa_2\in R^*$ where $a_1\in T^*$ and $a_2\in T$, by (3.7) and (3.8) we have

$$(\uppercase\expandafter{\romannumeral 1})_a=\sum_{\chi\in(R^{(s)^*}/G)^{\widehat{}}}{\chi}(a/\beta)G_{R^{(s)}}(\chi)\overline{G_R(\chi)}
$$
and
$$(\uppercase\expandafter{\romannumeral 2})_a=Q\sum_{\chi\in(\mathbb{F}_Q^*/\overline{D})^{\widehat{}}}
{\chi}(\overline{\alpha}_1/\overline{\beta}_1)G_Q(\chi)\overline{G_q(\chi})
$$

For $a=pa_2\in M\backslash\{0\}$ where $a_2\in T^*$, by (3.7) and (3.8) we have
\begin{align*}(\uppercase\expandafter{\romannumeral 1})_a&=\sum_{\chi\in(R^{(s)^*}/G)^{\widehat{}}}
{\chi}(a_2/\beta)G_{R^{(s)}}(\chi)\overline{G_R(\chi,\lambda_p)}\\&=q\sum_{\chi\in(R^{(s)^*}/G(1+M))^{\widehat{}}}
{\chi}(a_2/\beta)G_{R^{(s)}}(\chi)\overline{{G_q(\chi)}}
\end{align*}
and
\begin{align*}(\uppercase\expandafter{\romannumeral 2})_a&=Q\sum_{\chi\in(\mathbb{F}_Q^*/\overline{D})^{\widehat{}}}
{\overline{\chi}}(\overline{\beta}_1)G_Q(\chi)\overline{G_q(\chi,\lambda_0)}
\\&=Q(q-1)\sum_{\chi\in(\mathbb{F}_Q^*/\overline{D}\mathbb{F}_q^*)^{\widehat{}}}
{\overline{\chi}}(\overline{\beta}_1)G_Q(\chi)\\&=Q(q-1)\sum_{\chi\in (\mathbb{F}_Q^*/\langle^
{\overline{\xi}^{(s)^{e'}}}\rangle)^{\widehat{}}}{\overline{\chi}}(\overline{\beta}_1)G_Q(\chi)
\end{align*}
since $\overline{D}\mathbb{F}_q^*=\langle
{\overline{\xi}^{(s)^{e}}}\rangle\cdot\langle
{\overline{\xi}^{(s)^{\frac{Q-1}{q-1}}}}\rangle=\langle
{\overline{\xi}^{(s)^{e'}}}\rangle$, where $e'=gcd(e,\frac{Q-1}{q-1})$.

At last, for $a=0$ we have by (3.7) and (3.8) that
\begin{align*}(\uppercase\expandafter{\romannumeral 1})_a&=\sum_{\chi\in(R^{(s)^*}/G)^{\widehat{}}}
\overline{{\chi}}(\beta)G_{R^{(s)}}(\chi){G_R(\overline{\chi},1)}\\&=q(q-1)\sum_{\chi\in(R^{(s)^*}/GR^*)^{\widehat{}}}
\overline{{\chi}}(\beta)G_{R^{(s)}}(\chi)
\end{align*}
and
\begin{align*}(\uppercase\expandafter{\romannumeral 2})_a&=Q(q-1)\sum_{\chi\in (\mathbb{F}_Q^*/\langle^
{\overline{\xi}^{(s)^{e'}}}\rangle)^{\widehat{}}}{\overline{\chi}}(\overline{\beta}_1)G_Q(\chi)
\end{align*}
(3) From (2) we know that
\begin{eqnarray*}N_\beta(0)=\begin{cases}
\frac{n}{q}+\frac{n(q-1)}{q(Q-1)}\sum_{\chi\in (\mathbb{F}_Q^*/\langle^
{\overline{\xi}^{(s)^{e'}}}\rangle)^{\widehat{}}} \ \chi(1/\overline{b})G_Q(\chi)\leq\frac{n}{q}(1+(q-1)M_1), &if \  \beta=pb, b\in T^{(s)^*}\cr
\frac{n}{q^2}+\frac{n(q-1)}{q^2Q(Q-1)}[q\sum_{\chi\in (R^{(s)^*}/GR^*)^{\widehat{}}} \ \chi(\frac{1}{\beta})G_{R^{(s)}}(\chi)&if \ \beta=\beta_1+p\beta_2  \cr
+Q\sum_{\chi\in(\mathbb{F}_Q^*/\langle^
{\overline{\xi}^{(s)^{e'}}}\rangle)^{\widehat{}}} \ \chi(1/\overline{\beta}_1)G_Q(\chi)]\leq\frac{n}{q^2}(1+(q^2-1)M_2), &\beta_1\in T^{(s)^*}, \ \beta_2\in T^{(s)}\end{cases}\end{eqnarray*}
If $M_1< 1$ and $M_2< 1$, then $N_\beta(0)< n$ for all $\beta\in R^{(s)}\backslash\{0\}$ which means that $c_\beta(\beta\in R^{(s)})$ are distinct codewords in $C$. Therefore $|C|=|R^{(s)}|=Q^2$ and
 $$d_H(C)=\min \limits_{ \atop \beta\in R^{(s)}\backslash\{0\}}(n-N_\beta(0))=\min\{\frac{n(q-1)}{q}(1-M_1),\ \frac{N(q^2-1)}{q^2}(1-M_2)\}$$

This completes the proof of Theorem 3.1.\\

\noindent\textbf{Remark 3.2} (1) It is well-known that for any $\chi\in\widehat{\mathbb{F}}_Q^*$, we have $|G_Q(\chi)|=\sqrt{Q}$. Then by Lemma 2.1 we know that for $\chi\in \widehat{R^{(s)^*}}$,
$|G_{R^{(s)}}(\chi)|\leq Q$. Since $GR^*=\langle{\xi}^{(s)^{e'}}\rangle\times(1+pV)$ where $e'=gcd(e,\frac{Q-1}{q-1}), |V|=p^d$ where $0\leq d\leq rs \ (q=p^r, Q=q^s)$, the size of $GR^*$ is
$|GR^*|=|\langle{\xi}^{(s)^{e'}}\rangle|\cdot|V|=\frac{Q-1}{e'}\cdot p^d$. If both of sides $GR^*$ and $\langle{\xi}^{(s)^{e'}}\rangle$ are big enough such that
$$|GR^*|=\frac{Q-1}{e'}p^d>Q  \ \text{and} \ \frac{Q-1}{e'}>\sqrt{Q}$$
Then by the definition of $M_1$ and $M_2$,
$$M_1<\frac{\sqrt{Q}}{Q-1}|\mathbb{F}_Q^*/\langle{\overline{\xi}}^{(s)^{e'}}\rangle|=\frac{e'\sqrt{Q}}{Q-1}<1$$
$$M_2<\frac{q\cdot|R^{(s)^*}|\cdot Q}{(q+1)Q(Q-1)|GR^*|}+\frac{\sqrt{Q}e'}{(q+1)(Q-1)}=\frac{qQ}{(q+1)|GR^*|}+\frac{\sqrt{Q}e'}{(q+1)(Q-1)}$$
$$<\frac{q}{q+1}+\frac{1}{q+1}=1$$
 By Theorem 3.1(3), the size of $C$ is $|C|=Q^2$ and $d_H(C)=\min\{\frac{n(q-1)}{q}(1-M_1),\frac{n(q^2-1)}{q^2}(1-M_2)\}$.\\

 (2) For any $\gamma\in R^*$ and $\beta\in R^{(s)}$, $T_R^{R^{(s)}}(\gamma\beta)=\gamma T_R^{R^{(s)}}(\beta)$. Therefore $T_R^{R^{(s)}}(\gamma\beta)=0$ if and only if $T_R^{R^{(s)}}(\beta)=0$.
 We know that $G\cap R^*$ is a subgroup of $G$ with size $l=|G\cap R^*|$, and $G$ is a disjoint union of $n'$ cosets $a_j(G\cap R^*) \ (a_j\in G, 1\leq j \leq n')$
of the subgroup $G\cap R^*(\subseteq R^*)$, where $n'=n/l$. Now we construct the following $R-$linear code
\begin{align*}\widetilde{C}=\widetilde{C}(G)=\{\widetilde{c}_\beta=(T_R^{R^{(s)}}(\beta a_1),\cdots,T_R^{R^{(s)}}(\beta a_{n'}))\in R^{n'}:\beta\in R^{(s)}\}\tag{3.9}
\end{align*}
 We know that the parameters of $\widetilde{C}$ is $(n',|C|,d')_q$ where $n'=n/l$ and $d'=d(C)/l$.\\

 By Theorem 3.1, the values $N_\beta(a)$ and $d=d_H(C)$ are expressed in terms of
Gauss sums on Galois rings $R^{(s)},$ $ R$ and  finite fields $\mathbb{F}_Q$ and $\mathbb{F}_q.$ By Theorem 2.1, the values of  Gauss sums on Galois rings can be expressed by ones on finite fields. But in general case it is hard to calculate explicitly such values. Now we
consider several very special subgroup $G$ of $R^{(s)^*}$ for which we can get the values of $N_\beta(a)$ and $d_H(C)$ for $C=C(G)$.\\

For a $\mathbb{F}_p-$ subspace $A$ of $\mathbb{F}_Q$ , the dual subspace of $A$ is defined by
$$A^\perp=\{a\in \mathbb{F}_Q: T_p^Q(ax)=0 \ \text{for \ all} \ x\in A\}$$
This is a  $\mathbb{F}_p-$ subspace  of $\mathbb{F}_Q$ and dim$_{\mathbb{F}_p}A+$dim$_{\mathbb{F}_p}A^\perp=$dim$_{\mathbb{F}_p}\mathbb{F}_Q=rs$.
It is known that for any $a\in\mathbb{F}_Q,$
\begin{eqnarray*}\sum_{x\in A}\zeta_p^{T^Q_p(ax)}=\begin{cases}|A|, &if \ a\in A^\perp  \cr 0, &otherwise \end{cases}\end{eqnarray*}

\noindent\textbf{Theorem 3.3}(e=1 case) Assume that

\noindent(\romannumeral1) $q=p^r, Q'=q^{s'}, Q=Q'^p=q^s(s=ps'),$

\noindent \  \ \ \ $R=GR(p^2,r), \ R^{(s')}=GR(p^2,rs'),\  R^{(s)}=GR(p^2,rs).$

\noindent(\romannumeral2) $G=T^{(s)^*}\times(1+pV)$ and $\overline{V}^\perp\subseteq\mathbb{F}_{Q'}$, dim$_{\mathbb{F}_p}\overline{V}=d$
( From dim$\overline{V}^\perp\leq$dim${\mathbb{F}_{Q'}}=rs'$ and dim$\overline{V}^\perp=rs-$dim$\overline{V}=rs-d$ we
know that $rs\geq d\geq r(s-s')=r(p-1)s'$. )

Let $C=C(G)$ be the linear code over $R$ defined by (3.1). Then

\noindent(1) For $\beta=pb, \ b\in T^{(s)^*}$,
\begin{eqnarray*}N_\beta(a)=\begin{cases}0, &if \ a\in R^*  \cr Qp^dq^{-1}, &if \ a\in pT^*  \cr(Qq^{-1}-1)p^d, & if a=0\end{cases}\end{eqnarray*}
\noindent(2) For $\beta=\beta_1(1+p\beta_2)\in R^{(s)^*} \ (\beta_1\in T^{(s)^*},\beta_2\in T^{(s)})$,
\begin{eqnarray*}N_\beta(a)=\begin{cases}Qq^{-2}p^d, &if \ a\in R^* \ or \ "a\in pT^*, T^Q_{Q'}(\overline{\beta_2})+1 \notin S"
 \cr Qq^{-2}(p^d-q), &if \ a\in pT^* \ and \  T^Q_{Q'}(\overline{\beta_2})+1 \in S  \cr(Qq^{-2}-1)p^d+Qq^{-1}(q-1), & if a=0  \ and \ T^Q_{Q'}(\overline{\beta_2})+1 \in S
 \cr(Qq^{-2}-1)p^d, & if a=0 \ and \  T^Q_{Q'}(\overline{\beta_2})+1 \notin S \end{cases}\end{eqnarray*}
where $S$ is the dual of $\overline{V}^\perp(\subseteq\mathbb{F}_{Q'})$ in $\mathbb{F}_{Q'}$. Namely,
$$ S=\{a\in \mathbb{F}_{Q'}: T_p^{Q'}(ax)=0 \text{\ for \ all } \ x\in \overline{V}^\perp\}.$$
Particularly, $C$ has length $n=|G|=(Q-1)p^d,$ size $|C|=Q^2$ and minimum Hamming distance $d=d_H(C)=Qp^dq^{-1}(q-1).$\\
\noindent\textbf{Proof} \  It is obvious that $n=|G|=(Q-1)p^d.$\\
(1) can be derived from Theorem 3.1(1) directly. From $e=1$ we know that the summations in (3.3)
and (3.4) contain the trivial character $\chi=1$ only. Thus for $a\in pT^*$,
$$N_\beta(a)=\frac{n}{q}+\frac{n}{q(Q-1)}=Qq^{-1}p^d$$
$$N_\beta(0)=\frac{n}{q}-\frac{n(q-1)}{q(Q-1)}=(Qq^{-1}-1)p^d$$
and $N_\beta(a)=0$ for $a\in R^*$.\\
(2) Let $\beta=\beta_1(1+p\beta_2)\in R^{(s)^*}$ where $\beta_1 \in  T^{(s)^*}$ and $\beta_2\in T^{(s)}.$ By
 Theorem 3.1,
 $$N_\beta(a)=\frac{n}{q^2}+\frac{n}{q^2Q(Q-1)}((\uppercase\expandafter{\romannumeral1})_a+(\uppercase\expandafter{\romannumeral2})_a) \ \ \ (a\in R)$$

(2.1) If $a=a_1(1+pa_2)\in R^*$, where $a_1\in T^*$ and $a_2 \in T.$ By Theorem 3.1,\\
$$(\uppercase\expandafter{\romannumeral2})_a=Q\sum_{\chi\in(\mathbb{F}_Q^*/\mathbb{F}_Q^*)^{ \ \widehat{}}}\chi(\overline{a}/\overline{\beta}_1)G_Q(\chi)\overline{G_q(\chi)}=Q$$
\begin{align*}(\uppercase\expandafter{\romannumeral1})_a&=\sum_{\chi\in(R^{(s)^*}/G)^{ \ \widehat{}}}\chi({a}/{\beta})G_{R^{(s)}}(\chi)\overline{G_R({\chi})}\\
&=\sum_{\varphi_b\in(\frac{1+pT^{(s)^*}}{1+pV})^{ \ \widehat{}}}\varphi_b(\frac{1+pa_2}{1+p\beta_2})G_{R^{(s)}}(\varphi_b)\overline{G_R(\varphi_b)}\\
&=\sum\limits_{ b\in T^{(s)}\atop\overline{b}\in\overline{V}^\perp}\varphi_b(\frac{1+pa_2}{1+p\beta_2})G_{R^{(s)}}(\varphi_b)\overline{G_R(\varphi_b)}
\end{align*}
Recall that $\varphi_b(T^{(s)^*})=1$ and for $x\in T^{(s)}$, $\varphi_b(1+px)=\zeta_p^{T_p^Q(\overline{b}\overline{x})}$. If $x\in T$,
then $\overline{x}\in\mathbb{F}_q$ and $\varphi_b(1+px)=\zeta_p^{T^q_p(\overline{x}T^Q_q(\overline{b}))}$. Thus if
$T^Q_q(\overline{b})=0$ then $\varphi_b(1+pT)=1$ and $\varphi_b$  is trivial character of $R^*$ so that $G_R(\varphi_b)=0$ by Lemma 2.1.
On the other hand, from assumption (\romannumeral2) we know that $\overline{V}^\perp\subseteq\mathbb{F}_{Q'}$. If $\overline{b}\in\overline{V}^\perp$,
then $\overline{b}\in\mathbb{F}_{Q'}$ and $T^Q_q(\overline{b})=T^{Q'}_q(T^Q_{Q'}(\overline{b}))=T^{Q'}_q(p\overline{b})=0.$
 Therefore $(\uppercase\expandafter{\romannumeral1})_a=0$, and for all $a\in R^*,$
 $$N_\beta(a)=\frac{n}{q^2}+\frac{n}{q^2Q(Q-1)}(Q+0)=Qq^{-2}p^d$$
(2.2) For $a=pa_2, \ a_2\in T^*$, we have
$$(\uppercase\expandafter{\romannumeral2})_a=Q(q-1)\sum_{\chi=1}\chi(1/\overline{\beta}_1)G_Q(\chi)=-Q(q-1)$$
\begin{align*}(\uppercase\expandafter{\romannumeral1})_a&=-q\sum\limits_{ b\in T^{(s)^*}\atop\varphi_b\in(\frac{1+pT^{(s)}}{1+p(V+T)})^{\widehat{}}}\varphi_b(\frac{1}{1+p\beta_2})G_{R^{(s)}}(\varphi_b)\\
&=-q\sum\limits_{ b\in T^{(s)^*}\atop\overline{b}\in(\overline{V}+\overline{T})^\perp}\varphi_b(\frac{1}{1+p\beta_2})G_{R^{(s)}}(\varphi_b)
\end{align*}
From $\overline{V}^\perp\subseteq\mathbb{F}_{Q'}$ we know that $\overline{V}\supseteq\mathbb{F}_{Q'}^\perp$. From $s=ps'$ we get
$\mathbb{F}_{Q'}^\perp\supseteq\mathbb{F}_{Q'}.$ Therefore $\overline{V}\supseteq\mathbb{F}_{Q'}\supseteq\mathbb{F}_q=\overline{T}$
and $\overline{T}+\overline{V}=\overline{V}.$ Thus \\
$$(\uppercase\expandafter{\romannumeral1})_a=-q\sum\limits_{ b\in T^{(s)}\atop 0\neq\overline{b}\in\overline{V}^\perp}\overline{\zeta}_p^{T^Q_p(\overline{b\beta_2})}\zeta_{p^2}^{T^{R^{(s)}}_{Z_{p^2}}(b')}Q$$
by (2.4) where $b'=b$ for $p=2$ and $b'=-b$ for $p\geq3.$ From $\overline{b}\in\overline{V}^\perp\subseteq\mathbb{F}_{Q'}=\overline{T^{(s')}}$
we know that $b\in T^{(s')}\subseteq R^{(s')}$ and $b'\in R^{(s')}$. Therefore
$$T^{R^{(s)}}_{Z_{p^2}}(b')=T^{R^{(s')}}_{Z_{p^2}}(T^{R^{(s)}}_{R^{(s')}}b')=T^{R^{(s')}}_{Z_{p^2}}(pb')=pT^{R^{(s')}}_{Z_{p^2}}(b')$$
Then we get
\begin{eqnarray*}(\uppercase\expandafter{\romannumeral1})_a=-Qq\sum_{ 0\neq{b}\in\overline{V}^\perp}{\zeta}_p^{T^Q_p(b\overline{\beta_2})+T_p^{Q'}(b)}
=-Qq\sum_{ 0\neq{b}\in\overline{V}^\perp}{\zeta}_p^{T^{Q'}_p(b(T_{Q'}^Q(\overline{\beta}_2)+1)}\\
=\begin{cases}-qQ(|\overline{V}^\perp|-1)=qQ-qQ^2p^{-d}, &if \ T^Q_{Q'}(\overline{\beta}_2)+1\in S,  \cr qQ, &otherwise.\end{cases}\end{eqnarray*}
And for $a\in pT^*,$
\begin{eqnarray*}N_\beta(a)=\begin{cases}\frac{n}{q^2}+\frac{n}{q^2Q(Q-1)}(qQ-qQ^2p^{-d}-Qq+Q)=Qq^{-2}(p^d-q), &if \ T^Q_{Q'}(\overline{\beta}_2)+1\in S,
 \cr \frac{n}{q^2}+\frac{n}{q^2Q(Q-1)}(qQ-Qq+Q)=Qq^{-2}p^d, &otherwise.\end{cases}\end{eqnarray*}
(2.3) For $a=0$, $(\uppercase\expandafter{\romannumeral2})_0=-Q(q-1)$ and
\begin{eqnarray*}(\uppercase\expandafter{\romannumeral1})_0=q(q-1)\sum\limits_{ b\in T^{(s)^*}\atop\varphi_b\in(\frac{1+pT^{(s)}}{1+pV})^{ \ \widehat{}}}
\varphi_b(\frac{1}{1+p\beta_2})G_{R^{(s)}}(\varphi_b)
=\begin{cases}q(q-1)Q(|\overline{V}^\perp|-1), &if \ T^Q_{Q'}(\overline{\beta}_2)+1\in S,  \cr -q(q-1)Q, &otherwise.\end{cases}\end{eqnarray*}
Therefore
\begin{eqnarray*}N_\beta(0)=\begin{cases}\frac{n}{q^2}+\frac{n}{q^2Q(Q-1)}(q(q-1)Q(Qp^{-d}-1)-Q(q-1))\\=p^d(Qq^{-2}-1)+(q-1)Qq^{-1}, &if \ T^Q_{Q'}(\overline{\beta}_2)+1\in S,
 \cr \frac{n}{q^2}+\frac{n}{q^2Q(Q-1)}(-q(q-1)Q-(q-1)Q)=(Qq^{-2}-1)p^d, &otherwise.\end{cases}\end{eqnarray*}
From above computation we know that\\

 max$\{N_\beta(0): 0\neq\beta\in R^{(s)}\}$\\

 $=$max$\{p^d(Qq^{-1}-1),p^d(Qq^{-2}-1)+(q-1)Qq^{-1},(Qq^{-2}-1)p^d\}$\\

\noindent From $d\geq r(s-s')\geq r$ and by an elementary calculation we get
\begin{align*}p^d(Qq^{-2}-1)<p^d(Qq^{-2}-1)+(q-1)Qq^{-1}\leq(Qq^{-1}-1)p^d<n=(Q-1)p^d \tag{3.11}\end{align*}
which means that $N_\beta(0)<n$ for all $0\neq\beta\in R^{(s)}$. Namely,
$c_\beta \ (\beta\in R^{(s)})$ are distinct codewords of $C$. Therefore $|C|=|R^{(s)}|=Q^2$
and the minimum Hamming distance of $C$ is
$$d_H(C)=n-max\{N_\beta(0): 0\neq\beta\in R^{(s)}\}=p^d(Q-1)-p^d(Qq^{-1}-1)=p^dQq^{-1}(q-1).$$
This completes the proof of Theorem 3.3.\\

\noindent\textbf{Remark} (1) We have that $|S|=\frac{Q'}{|\overline{V}^\perp|}=\frac{Q'}{Q}|\overline{V}|=\frac{Q'p^d}{Q}.$
The number of $\overline{\beta}_2\in \overline{T^{(s)}}=\mathbb{F}_Q$ satisfying
$T^Q_{Q'}(\overline{\beta}_2)+1\in S$ is $|S|\cdot\frac{Q}{Q'}=p^d$. Theorem 3.3 gives the complete
Hamming weight distribution as shown in Table 1.

\begin{table}[!htbp]\footnotesize
\begin{tabular}{|c|c|c|c|c|c|c|}
\hline
\diagbox[width=5em,trim=l]{$\beta$}{$N_\beta(a)$}{$a$} &$a\in R^*$& $a\in pT^*$&$ a=0$&number of $\beta$\\
\hline
$\beta=\beta_1(1+p\beta_2)\atop\beta_1\in T^{(s)^*}, \ \beta_2\in T^{(s)} $
,$T^Q_{Q'}(\overline{\beta}_2)+1\in S$&{$Qq^{-2}p^d$ } &$Qq^{-2}(p^d-q)$ &$p^d(Qq^{-2}-1)+Qq^{-1}(q-1)$&$p^d(Q-1)$\\
\hline
$\beta=\beta_1(1+p\beta_2)\atop\beta_1\in T^{(s)^*}, \ \beta_2\in T^{(s)} $
$T^Q_{Q'}(\overline{\beta}_2)+1\notin S$ &{$Qq^{-2}p^d$ }&$Qq^{-2}p^d$ &$p^d(Qq^{-2}-1)$&$(Q-p^d)(Q-1)$\\
\hline
{$\beta\in pT^{(s)^*}$}&0&$Qq^{-1}p^d$&$(Qq^{-1}-1)p^d$&$Q-1$\\
\hline
{$\beta=0$}&0&0&$(Q-1)p^d$&1
\end{tabular}
\caption{ Complete Hamming weight distribution $N_\beta(a)$ of $C(G)$ in Theorem 3.3.}
\end{table}

(2) From $\overline{V}^\perp\subseteq\mathbb{F}_{Q'}$ we know that $\overline{V}\supseteq\mathbb{F}_{Q'}^\perp\supseteq\mathbb{F}_q$. Therefore $V\supseteq T$ and $G=T^{(s)^*}\times(1+pV)\supseteq T^*\times(1+pT)=R^*$. By using Remark 3.2(2), we
get the linear code $\widetilde{C}=\widetilde{C}(G)$ over $R$ defined by (3.9). The parameters of $\widetilde{C}$ is $(\widetilde{n},\widetilde{K},\widetilde{d})$, where
$\widetilde{n}=\frac{n}{|G\cap R^*|}=\frac{n}{|R^*|}=\frac{n}{q(q-1)}=\frac{(Q-1)p^d}{q(q-1)}$, $\widetilde{K}=|\widetilde{C}|=|C|=Q^2$ and $\widetilde{d}=d_H(\widetilde{C})=\frac{d}{q(q-1)}=p^dQq^{-2}$.\\

Next we consider case $e'=1$ where $e'=gcd(e,\frac{Q-1}{q-1}).$ From
$e|Q-1$ we know that $e'=1$ if and only if $e|q-1$ and $gcd(e,s)=1$ (where $q=p^r$ and $Q=q^s$).\\

\noindent\textbf{Theorem 3.4} Assume that condition (\romannumeral1) of Theorem 3.3 and the following (\romannumeral2) hold,\\
(\romannumeral2)$ \ G=\langle\xi^{(s)^e}\rangle\times(1+pV)$, where $T^{(s)^*}=\langle\xi^{(s)}\rangle$, $Q-1=ef$, $e'=gcd(e,\frac{Q-1}{q-1})=1$,
and $\overline{V}^\perp\subseteq\mathbb{F}_{Q'}$, dim$_{\mathbb{F}_p}\overline{V}=d.$

Let $C=C(G)$ and $\widetilde{C}=\widetilde{C}(G)$ be the linear codes over $R$ defined by
(3.1) and (3.9) respectively. Then the parameters of $C$ and $\widetilde{C}$ are
$(n,|C|,d)$ and $(\widetilde{n},\widetilde{|C|},\widetilde{d})$  respectively, where
$$n=|G|=\frac{(Q-1)p^d}{e}, \ \widetilde{n}=\widetilde{|G|}=\frac{e}{q(q-1)}n=\frac{(Q-1)p^d}{q(q-1)}.$$
$$d=d_H(C)=p^dQ(q-1)/eq, \ \widetilde{d}=d_H(\widetilde{C})=p^dQq^{-2}.$$

\noindent\textbf{Proof} The length of $C$ is $n=|G|=\frac{(Q-1)p^d}{e}$. From Theorem 3.1 and the calculation in the proof of Theorem 3.2 we get\\
(1) For $\beta=pb,\ b\in T^{(s)^*}$,
\begin{align*}N_\beta(0)&=\frac{n}{q}+\frac{n(q-1)}{q(Q-1)}\sum_{\chi\in(\mathbb{F}_Q^*/{\langle\overline{\xi}^{(s)^{e'}}\rangle})^{ \ \widehat{}}
}\chi(1/\overline{b})G_Q(\chi)\\
&=\frac{n}{q}-\frac{n(q-1)}{q(Q-1)}=(Qq^{-1}-1)p^d/e \ (\ since \ \langle\overline{\xi}^{(s)}\rangle=\mathbb{F}_Q^*  \ and \ e'=1).\end{align*}
(2) For $\beta=\beta_1(1+p\beta_2)\in R^{(s)^*} ,$ where $\beta_1\in T^{(s)^*},\beta_2\in T^{(s)}$,

\begin{equation*}N_\beta(0)=\frac{n}{q^2}+\frac{n}{q^2Q(Q-1)}
((\uppercase\expandafter{\romannumeral1})_0+(\uppercase\expandafter{\romannumeral2})_0) \tag{3.12}\end{equation*}
$$(\uppercase\expandafter{\romannumeral2})_0=Q(q-1)
\sum_{\chi\in(\mathbb{F}_Q^*/{\langle\overline{\xi}^{(s)^{e'}}\rangle})^{ \ \widehat{}}
}\chi(1/\overline{\beta}_1)G_Q(\chi)=-Q(q-1)$$
\begin{align*}(\uppercase\expandafter{\romannumeral1})_0&=q(q-1)
\sum_{\chi\in(R^{(s)^*}/GR^*)^{ \ \widehat{}}}\chi(1/{\beta})G_{R^{(s)}}(\chi)\\
&=\begin{cases}q(q-1)(|\overline{V}^\perp|-1)Q, &if \ T^Q_{Q'}(\overline{\beta}_2)+1\in S,  \cr -q(q-1)Q, &otherwise.\end{cases}
\end{align*}
Therefore, by (3.12) we get
\begin{eqnarray*}N_\beta(0)=\begin{cases}\frac{n}{q^2}+\frac{n}{q^2Q(Q-1)}(-Q(q-1)+q(q-1)Q(\frac{Q}{p^d}-1))=\frac{1}{e}[p^d(Qq^{-2}-1)+(q-1)Qq^{-1}], &if \ T^Q_{Q'}(\overline{\beta}_2)+1\in S,
 \cr \frac{n}{q^2}+\frac{n}{q^2Q(Q-1)}(-Q(q-1)-q(q-1)Q)=\frac{p^d(Qq^{-1}-1)}{e}, &otherwise.\end{cases}\end{eqnarray*}
From  inequalities (3.11) we know that $|C|=Q^2$ and
$$d=d_H(C)=n-p^d(Qq^{-1})/e=p^dQ(q-1)/eq.$$
Moreover, from $G=\langle\xi^{(s)^e}\rangle\times(1+pV)$, $R^*=\langle\xi^{(s)^{\frac{Q-1}{q-1}}}\rangle\times(1+pT)$ and
$V\supseteq T$, we get $G\cap R^*=\langle\xi^{(s)^t}\rangle\times(1+pT)$
where
$$t=lcm(e,\frac{Q-1}{q-1})=e\cdot\frac{Q-1}{q-1} \ \ (since \ e'=gcd(e,\frac{Q-1}{q-1})=1).$$
Therefore $l=|G\cap R^*|=\frac{q(Q-1)}{t}=\frac{q(q-1)}{e}.$ By remark 3.2(2),
we know that $\widetilde{n}=\frac{n}{l}=\frac{(Q-1)p^d}{q(q-1)},$ $|\widetilde{C}|=|C|=Q^2$ and $\widetilde{d}=d_H(\widetilde{C})=\frac{d}{l}=p^dQq^{-2}.$
This completes the proof of Theorem 3.4.\\

\noindent\textbf{{4. Homogeneous weight and Gray map}}\\

As a generalization of the Lee weight on $Z_4=GR(2^2,1) \ ([4])$, the homogeneous weight on Galois ring $R=GR(p^l,r)$ (and on more general finite chain rings) has been presented in [3,14,17]. Here we consider the $l=2$ case.\\

\noindent\textbf{Definition 4.1} The homogenous weight $w_{hom}: R=GR(p^2,r)\longrightarrow Z$ is
defined by
\begin{align*}w_{hom}(a)=\begin{cases}q-1, &if a\in R^*,  \cr q, &if a\in pT^*, \cr 0, &if a=0.\end{cases}
\end{align*}
where $q=p^r$. This mapping can be extended to ( for any $n\geq1$ )
$$w_{hom}: R^n\longrightarrow Z, \ \ w_{hom}(v)=\sum_{i=1}^{n}w_{hom}(v_i) \ \ ( \ for \ v=(v_1,\cdots,v_n)\in R^n)$$

If $C$ is a linear code over $R$ with length $n$, the minimum homogeneous distance of $C$ is defined by
\begin{align*}d_{hom}(C)&=min\{w_{hom}(c-c'): c,c'\in C, c\neq c'\}\\
&=min\{w_{hom}(c): 0\neq c\in C\}\end{align*}
The main motivation of using the homogeneous weight on $R^n$ is that we can find an isometric map
$\psi: (R^n,w_{hom}) \longrightarrow (\mathbb{F}_q^{nq},w_H)$ such that $\psi(C)$ is a good (non-linear) code
in $\mathbb{F}_q^{nq}$ for some codes $C$ in $R^n.$\\

\noindent\textbf{Definition 4.2} A mapping $\psi: R^n\longrightarrow \mathbb{F}_q^{nq}$ is called (generalized)
Gray map if $\psi$ is isometric. Namely, for $v,v' \in R^n$,
$$w_{hom}(v-v')=w_H(\psi(v)-\psi(v')).$$
Since isometric mapping should be injective, the following result can be derived from Definition 4.2 directly.\\

\noindent\textbf{Lemma 4.3} Let $C$ be a linear code over $R=GR(p^2,r)$ with length $n$,
$\psi: R^n \longrightarrow \mathbb{F}_q^{nq}$ be a Gray map. Then $\psi(C)$ is a (nonlinear) code over $\mathbb{F}_q$ with length $nq$, $|\psi(C)|=|C|$ and the minimum Hamming distance
$$d_H(\psi(C))=min\{w_H(v-v'): v,v'\in\psi(C), v\neq v'\}$$
is $d_{hom}(C)$.\\

Gray maps $\psi: R^n \longrightarrow \mathbb{F}_q^{nq}$ have been constructed in [3] and [17] by
algebraic and combinatorial ways respectively. Here we use the construction given in [3].\\

Let $\mathbb{F}_q=\{a_1,\cdots,a_q\}$. Each element $\beta$ of $R$ can be expressed uniquely by
$$\beta=\beta_0+p\beta_1 \ (\beta_0,\beta_1 \in T \ so \ that \ \overline{\beta}_0,\overline{\beta}_1\in \overline{T}=\mathbb{F}_q)$$
We define the mapping $\psi: R \longrightarrow \mathbb{F}_q^{q}$ by
\begin{align*}\psi(\beta)=&(a_1\overline{\beta}_0+
\overline{\beta}_1,a_2\overline{\beta}_0+\overline{\beta}_1,\cdots,a_q\overline{\beta}_0+\overline{\beta}_1)\\
=&(f(a_1),f(a_2),\cdots,f(a_q))\in\mathbb{F}_q^{q} \ \ (f(x)=\overline{\beta}_0x+\overline{\beta}_1 \in \mathbb{F}_q[x] )\tag{4.1}\end{align*}
The image of $\psi$ is
$$\psi(R)=\{(f(a_1),\cdots,f(a_q)): f(x)\in\mathbb{F}_q[x],\  degf(x)\leq1\}$$
which is the first-order generalized Reed-Muller code over $\mathbb{F}_q$.
From the well-known complete Hamming weight distribution of such RM code we can see
that the mapping $\psi: (R,w_{hom}) \longrightarrow (\mathbb{F}_q^{q},w_H)$ is isometric ([3], Theorem1.1). Then for each $n\geq1,$
$$\psi: (R^n,w_{hom}) \longrightarrow (\mathbb{F}_q^{nq},w_H), \ \psi(v)=(\psi(v_1),\cdots,\psi(v_n)) \ for \ v=(v_1,\cdots,v_n)\in R^n \ \ (4.2)$$
is a Gray map.\\

Now we compute the homogeneous weight of codewords in $C(G)$ and $\widetilde{C}(G)$ defined by (3.1) and (3.9) respectively.\\

\noindent\textbf{Theorem 4.4} Assume that\\
(\romannumeral1) \ $q=p^r, \ Q=q^s, \ R=GR(p^2,r)=T+pT, \ R^{(s)}=GR(p^2,rs)=T^{(s)}+pT^{(s)}.$\\
(\romannumeral2) \ $G=D\times(1+pV)$ be a subgroup of $R^{(s)}=T^{(s)^*}\times(1+pT^{(s)}),$ where
$T^{(s)^*}=\langle\xi^{(s)}\rangle$, $D=\langle\xi^{(s)^e}\rangle$, $Q-1=ef,$ dim$_{\mathbb{F}_q}\overline{V}=d.$

Let $C=C(G)$ and $\widetilde{C}=\widetilde{C}(G)$ be the linear codes over $R$ defined by (3.1) and (3.9)
 respectively. Then the homogeneous weight of $c_\beta\in C$ is
\begin{equation*}w_{hom}(c_\beta)=\begin{cases}
(q-1)n-\frac{n(q-1)}{Q(Q-1)}\sum_{\chi\in(R^{(s)^*}/GR^*)^{\widehat{}}
 \ }\chi(\frac{1}{\beta})G_{R^{(s)}}(\chi), &if \ \beta\in R^{(s)^*},
 \cr(q-1)n-\frac{n(q-1)}{Q-1}\sum_{\chi\in(\mathbb{F}_Q^*/\langle\overline{\xi}^{(s)^{e'}}\rangle)^{\widehat{}} \ }\chi(1/\overline{b})G_{Q}(\chi), &if \ \beta=pb, b\in T^{(s)^*},
 \cr 0, & if \ \beta=0.\tag{4.3}\end{cases}\end{equation*}
where $n=|G|=(Q-1)p^d/e$ is the length of $C$. Moreover, $w_{hom}(\widetilde{c_\beta})=w_{hom}(c_\beta)/l$
for each $\beta\in R^{(s)}$ where $l=|G\cap R^*|$.\\

\noindent\textbf{Proof } It is obvious that $w_{hom}(c_0)=0.$ For $0\neq\beta\in R^{(s)}$, by definition we have
\begin{align*}w_{hom}(c_\beta)&=\sum_{a\in R}N_\beta(a)w_{hom}(a)=(q-1)\sum_{a\in R^*}N_\beta(a)+q\sum_{a\in pT^*}N_\beta(a)\\&=(q-1)\sum_{a\in R}N_\beta(a)+\sum_{a\in pT^*}N_\beta(a)-(q-1)N_\beta(0)\\&=(q-1)n
+\sum_{a\in pT^*}N_\beta(a)-(q-1)N_\beta(0)\tag{4.4}
\end{align*}
(A). For $\beta=pb,$ $b\in T^{(s)^*}$, from (3.4) and (3.3) we have
$$N_\beta(0)=\frac{n}{q}+\frac{n(q-1)}{q(Q-1)}\sum_{\chi\in(\mathbb{F}_Q^*/\langle\overline{\xi}^{(s)^{e'}}\rangle)^{\widehat{}} \ }\chi(1/\overline{b})G_{Q}(\chi)$$
\begin{align*}
\sum_{a_2\in T^*}N_\beta(pa_2)&=\frac{n(q-1)}{q}+\frac{n}{q(Q-1)}\sum_{\chi\in(\mathbb{F}_Q^*/\langle\overline{\xi}^{(s)^{e'}}\rangle)^{\widehat{}} \ }\chi(1/\overline{b})G_{Q}(\chi)\overline{G_q(\chi)}\sum_{a_2\in T^*}\chi(\overline{a}_2)
\\
&=\frac{n(q-1)}{q}-\frac{n(q-1)}{q(Q-1)}\sum_{\chi\in(\mathbb{F}_Q^*/\langle\overline{\xi}^{(s)^{e'}}\rangle)^{\widehat{}} \ }\chi(1/\overline{b})G_{Q}(\chi)
\end{align*}
Then from (4.4) we get
$$w_{hom}(c_\beta)=(q-1)n-\frac{(q-1)n}{Q-1}\sum_{\chi\in(\mathbb{F}_Q^*/\langle\overline{\xi}^{(s)^{e'}}\rangle)^{\widehat{}} \ }\chi(1/\overline{b})G_{Q}(\chi)$$
(B). For $\beta=\beta_1(1+p\beta_2)\in R^*$ where $\beta_1\in T^{(s)^*}$ and $\beta_2\in T^{(s)}.$  From Theorem 3.1 we get
$$N_\beta(0)=\frac{n}{q^2}+\frac{n}{q^2Q(Q-1)}[q(q-1)\sum_{\chi\in(R^{(s)^*}/GR^*)^{\widehat{}}
 \ }\chi(\frac{1}{\beta})G_{R^{(s)}}(\chi)+Q(q-1)
 \sum_{\chi\in(\mathbb{F}_Q^*/\langle\overline{\xi}^{(s)^{e'}}\rangle)^{\widehat{}} \ }\chi(1/\overline{\beta}_1)G_{Q}(\chi)]$$
$$\sum_{a=pa_2\in pT^*}N_\beta(a)=\frac{n(q-1)}{q^2}+\frac{n}{q^2Q(Q-1)}\sum_{a=pa_2\in pT^*}((\uppercase\expandafter{\romannumeral1})_a+(\uppercase\expandafter{\romannumeral2})_a)$$
where
\begin{align*}\sum_{a=pa_2\in pT^*}(\uppercase\expandafter{\romannumeral1})_a&=q\sum_{\chi\in(R^{(s)^*}/G(1+M))^{ \ \widehat{}}}
\chi(1/\beta)G_{R^{(s)}}(\chi){G_q(\chi)}\sum_{a\in T^*}\chi(a_2)\\
&=-q(q-1)\sum_{\chi\in(R^{(s)^*}/GR^*)^{ \ \widehat{}}}
\chi(1/\beta)G_{R^{(s)}}(\chi)
\end{align*}
$$\sum_{a\in pT^*}(\uppercase\expandafter{\romannumeral2})_a
=Q(q-1)^2\sum_{\chi\in(\mathbb{F}_Q^*/\langle\overline{\xi}^{(s)^{e'}}\rangle)^{\widehat{}}}\chi(1/\overline{\beta}_1)G_Q(\chi)$$
Then from (4.4) we get
$$w_{hom}(c_\beta)=(q-1)n-\frac{n}{Q(Q-1)}(q-1)\sum_{\chi\in(R^{(s)^*}/GR^*)^{ \ \widehat{}}}\chi(1/\beta)G_{R^{(s)}}(\chi)$$
This completes the proof of (4.3). Moreover, $aR^*=R^*$ and $a(pT^*)=pT^*$ for any $a\in R^*.$
This implies that for each $a\in G\cap R^*$ and $x\in G$, $T_R^{R^{(s)}}(\beta ax)=T_R^{R^{(s)}}(\beta x)$
so that $w_{hom}(c_\beta)=w_{hom}(c_{a\beta})$. Therefore for each codeword $\widetilde{c_\beta}\in \widetilde{C}(G)$, $w_{hom}(\widetilde{c_\beta})=w_{hom}(c_\beta)/l.$ This completes the proof of Theorem 4.4.\\

Applying Theorem 4.4 to $C(G)$ and $\widetilde{C}(G)$ where $G$ is given in Theorem 3.4, we get the following result.\\

\noindent\textbf{Theorem 4.5} ($e'=1$ case) Assume that\\
(\romannumeral1) \ $q=p^r, \ Q'=q^{s'}, Q=Q'^p=q^s \ (s=ps')$\\
 $R=GR(p^2,r), R^{(s')}=GR(p^2,rs'),R^{(s)}=GR(p^2,rs)$\\
(\romannumeral2) \ $G=\langle\xi^{(s)^e}\rangle\times(1+pV)$ , where
$T^{(s)^*}=\langle\xi^{(s)}\rangle$,  $Q-1=ef,$ and $e'=gcd(e,\frac{Q-1}{q-1})=1$.
Then the homogeneous weight distribution of the linear codes $C=C(G)$ and $\widetilde{C}=\widetilde{C}(G)$ over $R$ are shown in Table 2.

\begin{table}[!htbp]\footnotesize
\begin{tabular}{|c|c|c|c|c|}
\hline
\multicolumn{2}{|c|}{$\beta$}& $w_{hom}(c_\beta)$& $w_{hom}(\widetilde{c_\beta})$ &number of $\beta$\\
\hline
\multirow{2}{*}{$\beta\in R^{(s)^*}, \ \beta=\beta_1(1+p\beta_2)\atop\beta_1\in T^{(s)^*}, \ \beta_2\in T^{(s)} $ }
&$T^Q_{Q'}(\overline{\beta}_2)+1\in S$ &$Q(q-1)(p^d-1)/e$ &$Q(p^d-1)q^{-1}$&$p^d(Q-1)$\\
 &$T^Q_{Q'}(\overline{\beta}_2)+1\notin S$ &$Q(q-1)p^d/e$&$Qq^{-1}p^d$ &$(Q-p^d)(Q-1)$\\
\hline
\multicolumn{2}{|c|}{$\beta\in pT^{(s)^*}$}&$Q(q-1)p^d/e$&$Qq^{-1}p^d$&$Q-1$\\
\hline
\multicolumn{2}{|c|}{$\beta=0$}&0&0&1
\end{tabular}
\caption{ Homogeneous weight distribution of $C(G)$ and $\widetilde{C}(G)$}
\end{table}
\noindent where $S$ is defined in Theorem 3.4.\\

\noindent\textbf{Proof} \ The length of $C(G)$ is $n=|G|=\frac{Q-1}{e}p^d$.\\
(A). For $\beta=pb,$ $b\in T^{(s)^*}$, from (4.1) and $e'=1$ we get
$$w_{hom}(c_\beta)=(q-1)n+\frac{(q-1)n}{Q-1}=\frac{(q-1)nQ}{Q-1}=\frac{Q(q-1)p^d}{e}$$
(B.) For $\beta=\beta_1(1+p\beta_2)\in R^{(s)*}$ , from (4.1) and  $e'=1$ we get
$$w_{hom}(c_\beta)=(q-1)n-\frac{(q-1)n}{Q(Q-1)}\sum\limits_{b\in T^{(s)^*}\atop \overline{b}\in \overline{V}^\perp}\varphi_b(1/1+p\beta_2)G_{R^{(s)}}(\varphi_b)$$

We have showed in the proof of Theorem 3.3 that the summation of the right-hand side is $Q(|\overline{V}^\perp|-1)$ if $T^Q_{Q'}(\overline{\beta}_2)+1\in S$, or $-Q$ otherwise where
$|\overline{V}^\perp|=Qp^{-d}$. Therefore
\begin{equation*}w_{hom}(c_\beta)=(q-1)n+\begin{cases}
\frac{-n(q-1)(Qp^{-d}-1)}{Q-1}
 \cr\frac{n(q-1)}{Q-1}\end{cases}
 =\begin{cases}
\frac{Q(q-1)(p^{d}-1)}{e}, &\ if \ T^Q_{Q'}(\overline{\beta}_2)+1\in S,
 \cr\frac{Q(q-1)p^d}{e}, & \ otherwise.\end{cases}
 \end{equation*}
This completes the computation of $w_{hom}(c_\beta)$ as shown in Table 2.
At last, we have $w_{hom}(\overline{c}_\beta)=w_{hom}(c_\beta)/l$, where
$l=|G\cap R^*|=\frac{q(q-1)}{e}.$ This gives the values of $w_{hom}(\widetilde{c_\beta})$.\\

Finally, we use the Gray map to obtain series of nonlinear codes over $\mathbb{F}_q.$
The following result can be derived from Lemma 4.3 directly.\\

\noindent\textbf{Corollary 4.6} Let $C=C(G)$ and  $\widetilde{C}=\widetilde{C}(G)$ be the
linear codes over $R$ given in Theorem 4.5, $\psi$ be the Gray map defined by (4.1) and (4.2).
Then $\psi(C)$ and $\psi(\widetilde{C})$ are codes over $\mathbb{F}_q$ with length $qn=\frac{(Q-1)qp^d}{e}$
and $\frac{(Q-1)p^d}{q-1}$ respectively. $|\psi(C)|=|\psi(\widetilde{C})|=Q^2.$ The Hamming weight distribution of $\psi(C)$ and $\psi(\widetilde{C})$ are the same as shown in Table 2, but instead of
$w_{hom}(c_\beta)$ and $w_{hom}(\widetilde{c_\beta})$ by $w_H(\psi(c_\beta))$ and $w_H(\psi(\widetilde{c_\beta}))$ respectively.  Particularly, both of $\psi(C)$ and $\psi(C')$
are two-distance code. Their minimum Hamming distance are
$$d_H(\psi(C))=Q(q-1)(p^d-1)/e, \ d_H(\psi(\widetilde{C}))=Q(p^d-1)q^{-1}.$$

\noindent\textbf{{5. Conclusion}}\\

In this paper we consider a series of linear codes $C(G)$ and $\widetilde{C}(G)$ over Galois ring
$R=GR(p^2,r)$ where $G=D\times(1+pV)$ is a subgroup of $R^{(s)^*}=T^{(s)^*}\times(1+pT^{(s)})$
$(R^{(s)}=GR(p^2,rs))$. We have presented a general formula on $N_\beta(a)$ in terms of Gauss sums on
$R^{(s)}$ where $N_\beta(a)$ is the number of a-component of the codeword $c_\beta=(T_R^{R^{(s)}}(\beta x))_{x\in G}$ in $C(G)$ for $a\in R$ and $\beta\in R^{(s)}$ (Theorem 3.1). We have determined the complete
Hamming weight distribution of $C(G)$ and the minimum Hamming distance of $\widetilde{C}(G)$ if $D=T^{(s)^*}$ and $1+pV$ is some special subgroup of $1+pT^{(s)}$ (Theorem 3.3). If $e'=[T^{(s)^*}:GR]=1,$
we have determined the minimum Hamming distance of $C(G)$ and $\widetilde{C}(G)$ (Theorem 3.4).
Then we have introduced the homogeneous weight $w_{hom}$ on $R^n$ and showed a general formula
$w_{hom}(c_\beta)$, $w_{hom}(\widetilde{c_\beta})$ for codeword $c_\beta\in C(G)$ and
$\widetilde{c_\beta}\in \widetilde{C}(G)$ (Theorem 4.4). For the special subgroup $G$ in Theorem 3.4, we
have determined the homogeneous weight distribution of $C(G)$ and $\widetilde{C}(G)$ (Theorem 4.5).
Finally we  obtained series of nonlinear codes over $\mathbb{F}_q$ with two Hamming distances by
using Gray map (Corollary 4.6).

Our consideration is mainly concerned on case $e'=1$ in this paper. If $e'\geq2$ more computation on
Gauss sums on $R$ will be involved. In consequent paper we will deal with some case of  $e'\geq2$ where
the Gauss sums can be determined explicitly. Moreover, Theorem 3.1 can be viewed as a generalization of
irreducible cyclic codes over finite fields, as we indicated in section 1. Recently, several techniques have been found to determine the Hamming weight distribution of reducible cyclic codes over finite fields.
It will be interesting to know if such techniques can be used in Galois ring case.

\dse{~~References}
\noindent [1] L.D.Baumert, R.J.McEliece, Weights of irreducible cyclic codes, Information and Control, 20(1972), 158-175.\\

\noindent [2] B.C.Berndt, R.J.Evans and K.S.Williams, \emph{Gauss and Jacobi Sums}, Wiley-Interscience Pub., New York, 1998.\\

\noindent [3]M.Gregerath and S.E.Schmidt, Gray isometries for chain rings and a nonlinear ternary $(36,3^{12},5)$ code, IEEE Trans. Inform. Theory, 20(7), 1999, 2522-2524.\\

\noindent[4] A.R. Hammons Jr., P. V. Kumar, A.R. Calderbank, N.J.A. Sloane and P.
Sol$\acute{\text{e}}$, The $\mathbb{Z}_4$-linearity of Kerdock,
Preparata, Goethals, and related codes, IEEE Trans. Inform. Theory,
1994, 40: 301-319.\\

\noindent [5] H.M.Kiah, K.H.Leung and S.Ling, cylcic codes over $GR(P^2,m)$
of length $p^k$, Finite Fields and Applic. 14(2008), 834-846.\\

\noindent [6] M.K. Han, K.H. Leung, S. Ling, A note on cyclic codes over $GR(p^2,m)$ of length $p^k$, Des. Codes, Cryptogr. 63(1), 2012, 105-112.\\

\noindent[7] T.R. Kwon and W.S. Yoo, Remarks on Gauss sums over
Galois rings, Korean J. Math., 2009, 17: 43-52.\\

\noindent[8] J.E.Lamprecht, Calculation of general Gauss sums and quadratic Gauss sums in finite rings, Theorie des Nombres(Quebec, 1987), de Gruyter, Berlin, 1989, 561-573.\\

\noindent[9] P.Langevin, A new class of two weight codes,  Finite Fields and Their Applications (Glasgow, 1995),S.Cohen and H. Niederreiter, eds. Cambridge Univ. Press, Cambridge, 1996, 181-187.\\

\noindent[10] Jin Li, Shixin Zhu, Keqin Feng, The Gauss sums and Jacobi sums over Galois ring $GR(p^2,r),$ Science China, 56(2013), 1457-1465.\\

\noindent [11] Lidl and Niederreiter, \emph{Finite Fields}, Addison-Wesley, London, 1983.\\

\noindent [12] R.J.McEliece, Irreducible cyclic codes and Gauss sums, Combinatorics,(M.Hall and Van Lint eds.) Reidel, Dordrecht-Boston,1975: 185-202.\\

\noindent[13] Y. Oh and H.J.
Oh, Gauss sums over Galois rings of characteristic 4,
Kangweon-Kyungki Math. Jour., 2011, 9: 1-7.\\

\noindent[14] J.F. Voloch and J.L.Walker, Homogeneous weights and exponential sums, Finite Fields and Their Applications, 9(2003), 310-321.\\

\noindent[15]  ZheXian Wan, Cyclic codes over Galois rings, Alg.Colloq.,6(1999),291-304.\\

\noindent[16]  ZheXian Wan, \emph{Lecture Notes on Finite Fields and Galois Rings}, Word
Scientific, Singapore, 2003.\\

\noindent [17] B. Yildiz, A combinatorial construction of the Gray map over Galois rings, Disc. Math.
309(2009), 3408-3412.\\

\noindent [18] M. Van Der Vlugt, Hasse-Davenport curve, Gauss sums and weight distribution
of irreducible cyclic codes, J. Number Theory, 55(1995), 145-159.\\

\noindent [19] P. Sol$\acute{e}$, V. Sison, Bounds on the Minimum Homogeneous Distance
of the $p^r$-ary Image of Linear Block Codes
over the Galois Ring $GR(p^r, m)$,  IEEE Trans. Inform. Theory, 53(2007), 2270-2273.\\

\noindent [20] M. Bhaintwal, Skew quasi-cyclic codes over Galois rings, Designs Codes and Cryptography, 62(2012), 85-101.
\end{document}